\documentclass[11pt]{article}

\usepackage[preprint]{acl}

\usepackage{times}
\usepackage{latexsym}
\usepackage{url}

\usepackage[T1]{fontenc}
\usepackage{amsmath}
\usepackage{booktabs}
\usepackage{array}
\usepackage{tabularx}
\usepackage{ragged2e}
\usepackage{hyperref}
\usepackage{float}

\usepackage[utf8]{inputenc}

\usepackage{microtype}

\usepackage{inconsolata}

\usepackage{graphicx}

\newcolumntype{Y}{>{\RaggedRight\arraybackslash}X}

\title{Humans Introduce, Models Elaborate: Asymmetric Narrative Agency in Human--LLM Co-Writing}

\author{
  \textbf{Halfdan Nordahl Fundal\textsuperscript{1,2}},
  \textbf{Yuri Bizzoni\textsuperscript{1,3}},
  \textbf{Charlotte Gjørup Bilde\textsuperscript{1,2}}, \\
  \textbf{Ida Bække Johannesen\textsuperscript{1,2}},
  \textbf{Rebekah Baglini\textsuperscript{1,2}} \\
  \textsuperscript{1}TEXT: Center for Contemporary Cultures of Text, Aarhus University,\\
  \textsuperscript{2} Department of Linguistics, Cognitive Science, and Semiotics, Aarhus University,\\
  \textsuperscript{3} Center for Humanities Computing, Aarhus University
\\
  \small{
    \textbf{Correspondence:} \href{mailto:halfi@cc.au.dk}{halfi@cc.au.dk}
  }
}

\begin{document}
\maketitle

\begin{abstract}
Human--LLM co-writing is increasingly used for open-ended text generation, but much prior work focuses on final outputs rather than the interactional dynamics through which stories are produced. We study turn-based collaborative storytelling across three matched conditions: Human--Human (HH), Human--LLM (HA), and LLM--LLM (AA). Using a shared storytelling paradigm, we measure how agents align, introduce novel material, and influence narrative development through turn-level measures of valence adaptation, semantic novelty, transience, and resonance. Our results show that HA co-writing is not intermediate between HH and AA collaboration. Instead, it displays a distinctive asymmetry where humans tend to introduce more novel and persistent narrative material, while LLMs tend to elaborate and stabilize the existing context. These findings suggest that, in this setting, LLMs function less as human co-authors and more as adaptive narrative amplifiers that reshape how agency is distributed in collaborative writing.
\end{abstract}

\section{Introduction}
Large language models (LLMs) are increasingly used as partners in open-ended writing tasks, from ideation and revision to full narrative co-creation \citep{noy2023experimental,clark-smith-2021-choose,cheng2025inspiration}. In such settings, the relevant question is how the text came to be. A co-written story unfolds through a sequence of turns: each contribution responds to prior context, shifts the semantic and emotional trajectory of the story, and constrains what can come next. Understanding Human--LLM co-writing therefore requires methods that model collaboration as a dynamic process.
Turn-based storytelling provides a controlled setting for studying this process. 
Existing work on Human--LLM co-creation has focused on output quality, creativity, productivity and user experience. These perspectives are important, but leave open an interactional question: does Human--LLM co-writing resemble Human--Human collaboration, LLM--LLM collaboration, or a distinct regime of narrative coordination? Answering this question requires direct comparison of different dyadic pairings on the same experimental paradigm.
\begin{figure}[t]
      \centering
  \includegraphics[width=\linewidth]{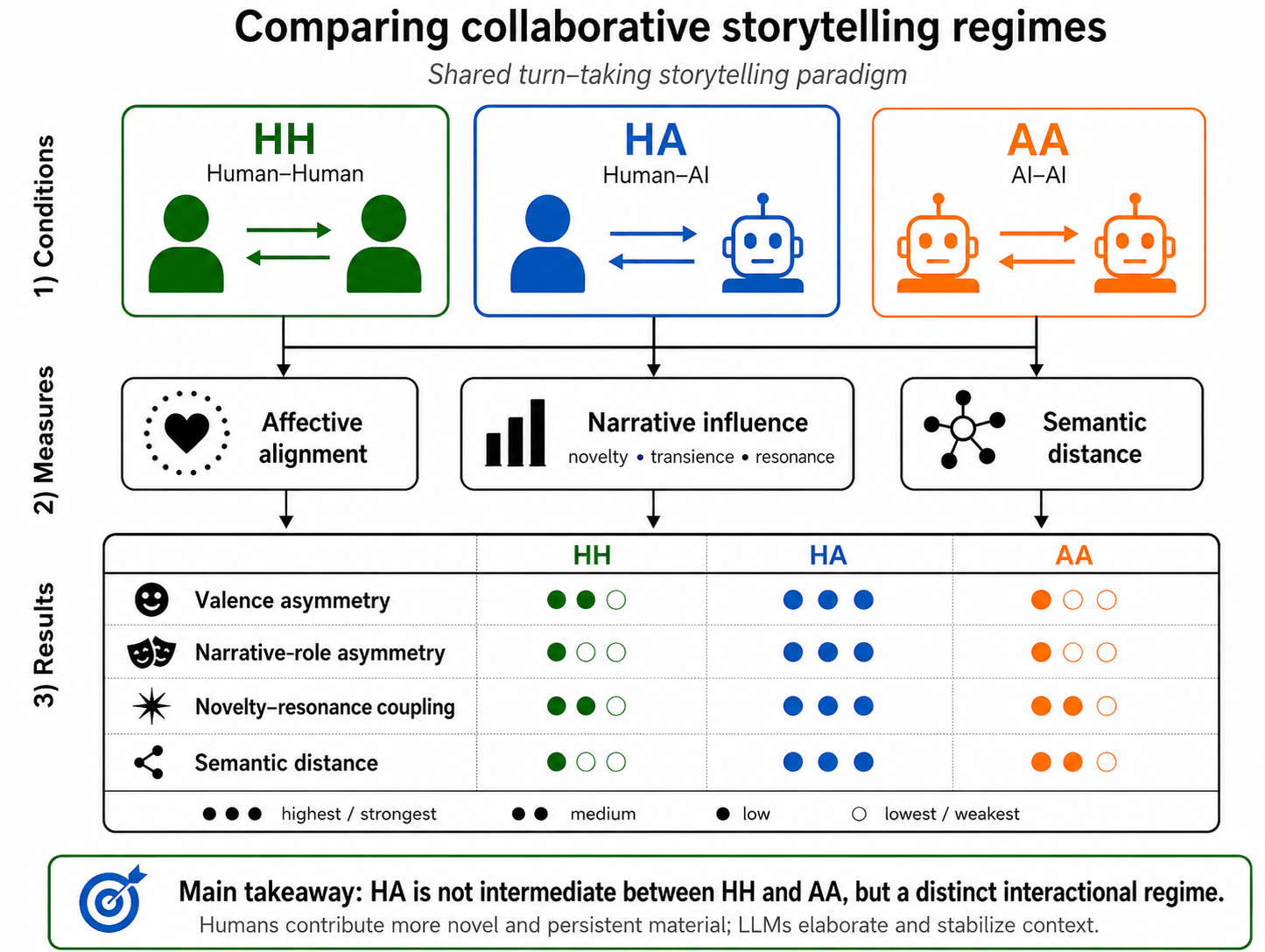}
    \caption{\textbf{Overview of setup and pipeline.} Stories are generated under HH, HA, and AA conditions, then analyzed for valence alignment, narrative influence, and semantic distance.}
    \label{fig:overview}
\end{figure}
Figure~\ref{fig:overview} summarizes the shared setup and analysis pipeline.
We address this gap by comparing three turn-based co-writing conditions: \textbf{Human--Human (HH), Human--LLM (HA)}, and \textbf{LLM--LLM (AA)}. Building on an existing Human--LLM corpus, we introduce matched Human--Human and LLM--LLM conditions, allowing us to compare collaborative dynamics while holding the task structure constant. We analyze each story at the turn level using measures of affective alignment and semantic influence. 
This comparative design lets us test whether asymmetry is a general property of dyadic storytelling or a specific feature of HA collaboration. We treat narrative agency as a measurable distribution of local influence, addressing who shifts the story, whose contributions are taken up, and how strongly partners adapt to each other.

We investigate three research questions:

\begin{enumerate}
    \item \textbf{RQ1:} How do affective alignment and semantic distance differ across HH, HA, and AA co-writing?
    \item \textbf{RQ2:} How does narrative influence distribute across HH, HA, and AA co-writing?
    \item \textbf{RQ3:} Does HA constitute a distinct interactional regime or an intermediate condition between HH and AA?
\end{enumerate}

Our contributions are threefold. First, we construct a matched comparative dataset of HH, HA, and AA turn-based stories. Second, we apply a turn-level framework for measuring affective adaptation and directional narrative influence in collaborative writing. Third, we show that Human--LLM co-writing exhibits its clearest asymmetry in semantic influence: humans tend to contribute more novel and persistent material, while LLMs tend to elaborate the evolving narrative context. This suggests that LLMs do not simply replace human authors, but redistribute narrative agency in a unique way.

\section{Related Work}

\paragraph{Human--LLM co-creation}
Human--LLM co-writing holds the promise of hybrid intelligence: the possibility that human and artificial writers may collaborate to produce new forms of creativity \cite{cheng2025inspiration, rafner2023creativity}. Much prior work on human--LLM creativity evaluates output and task-level outcomes such as quality, originality, productivity, or ideation, with mixed findings. \citet{Domanti_2026} compared human and LLM creativity by relating semantic network diversity to originality, showing that LLM-generated outputs may be more original than outputs of low-creative humans. In specific convergent and divergent thinking tasks, LLMs have been shown to outperform student participants \cite{arora2025generative}. Human--LLM co-creation may furthermore increase the absolute number of novel artifacts \cite{zhou2025expands}, task enjoyment, and writing quality \cite{noy2023experimental}. On the other hand, co-creation can narrow content diversity, homogenizing ideas and style \cite{wenger2025weredifferentweresame,anderson2024homogenization}. Recent work suggests that human--LLM joint creativity may reduce participants' self-perceived creativity \cite{luchini2025creative} and, in some settings, may underperform human--human collaboration on creative outcomes \cite{tang2025best}.

\paragraph{Alignment and affective adaptation}
A second line of work studies alignment and adaptation in dialogue, showing that human and artificial agents exhibit forms of linguistic and emotional coordination. Such coordination is observable on semantic, affective, syntactic, phonetic and lexical levels, and can support effective communication \cite{branigan2010linguistic}. \citet{koulouri2016and} showed how lexical alignment occurs in human--computer exchanges, through vocabulary adaptation. Other lexical approaches have revealed differences between human--human and human--agent dialogue \cite{duplessis2021towards, li2025ai}, and have shown consistent human alignment to artificial agents \cite{ostrand2023rapid}. However, lexical matching captures one limited dimension of coordination, and may overlook the underlying emotional contagion that unfolds during dialogue \cite{hatfield1993emotional}. On this account, \citet{varni2017computational} proposed a computational framework for assessing emotional contagion by matching time series polarity states between interlocutors, while other approaches have modeled it by categorizing utterances into predefined emotion labels \cite{poria2019emotion}. 

\paragraph{Sentiment analysis in digital humanities}
For modeling affective dynamics, computational sentiment analysis has emerged as a valuable methodological foundation \cite{mohammad2016sentiment}. Prior work has modeled emotional arcs and sentiment progression in literary narratives \citep{reagan2016emotional, hu2021dynamic, ohman2022computational}. However, dictionary-based approaches have weak context awareness, motivating the use of transformer-based alternatives \cite{feldkamp2024comparing}. Both methods remain challenging to apply as general-purpose sentiment models to literary text. \citet{Lyngbaek_2026} addressed this by using concept vector projection as a method for deriving continuous sentiment scores across domains and languages. Building on this, \citet{fundal2026directionalalignmentnarrativeagency} adopted concept vector projection to model sentiment alignment between human--LLM interlocutors in co-creative writing.  

\paragraph{Information propagation in static discourse and dialogue}
A central question in discourse analysis concerns how contributions diverge from prior context and whether they shape what follows. \citet{barron2018individuals} introduced an information-theoretic framework for measuring novelty, transience, and resonance in large-scale corpora of French Revolution speeches. In their paradigm, novelty captures departure from prior discourse, transience captures limited future uptake, and resonance encompasses contributions that are both novel and persistent. Extending these information-theoretic intuitions to conversational paradigms, \citet{bergey2024yeah} used LLM raw surprisal scores of utterances to model information flow and to predict backchannels in dialogue, and \citet{maes-etal-2022-shared} used a similar approach to show how the stability of surprisal-based information varies locally within themes. Other work on dialogue has shown how information density converges between interlocutors over time \cite{xu2018information}. However, how these information-propagation measures operate in co-creative human--LLM narratives remains less explored.



\section{Datasets}


\begin{table}[t]
\centering
\small
\setlength{\tabcolsep}{3pt}
\begin{tabular}{@{}lrrrr@{}}
\toprule
Condition & Stories & Exchanges & Turns & Mean words \\
\midrule
HH & 36 & 360 & 720 & 26.13 \\
HA & 97 & 873 & 1746 & 27.28 \\
AA & 80 & 800 & 1600 & 28.24 \\
\bottomrule
\end{tabular}
\caption{Dataset statistics across co-writing conditions. \textit{Turns} are individual agent contributions; \textit{exchanges} are adjacent pairs we use for alignment and influence analyses. Counts describe the final analytic corpus; individual analyses use subsets of it, since turn-level models drop turns with missing scores (Table~\ref{tab:valence-descriptives}) and transience is undefined for the final exchange of each story.}
\label{tab:dataset}
\end{table}

We use a turn-based storytelling paradigm across three dyadic conditions to compare collaborative dynamics while holding a constant task structure\footnote{Datasets and analysis code: \url{https://github.com/Halfidaldal/penpal-emnlp-acl}}. 
Participant recruitment, preprocessing, and demographics are in Appendix~\ref{sec:appendix}.

\subsection{Shared task design}
Across conditions, agents received identical symmetric instructions:
\textit{“You are an author taking part in a collaborative storytelling activity with another author. Together, you will create a story by taking turns adding to it. Your goal is to continue from where your partner has left off... You have 10 exchanges to write the story.”} 

We designed the prompt to elicit open-ended narrative continuation. The fixed turn-taking format makes each story analyzable as a sequence of local responses and allows us to measure how each agent adapts to the preceding contributions.

Because the prompt is identical across HH, HA, and AA, any role specialization in HA reflects dyadic interaction rather than instruction bias. We analyze individual \textbf{turns} and adjacent \textbf{directed exchanges} to measure response, alignment, and influence.

\paragraph{Human--LLM}
The Human--LLM data comes from \citet{fundal2026directionalalignmentnarrativeagency}. In this condition, human participants alternated turns with an LLM in the collaborative storytelling interface. There were 97 participants (\textit{M}$_{\text{age}}$ = 27.34 years, \textit{SD} = 14.66); of those who reported gender, 34 were male, 40 female, and 4 other. 
The LLMs used in this condition and in our AA condition were \path{gpt-4.1-2025-04-14}, \path{claude-sonnet-4-5-20250929}, \path{Llama-3.3-70B-Instruct}, and \path{Qwen2.5-72B-Instruct}.

\paragraph{Human--Human} 
The HH condition used the same turn-taking structure as the HA condition and included 72 participants (\textit{M}$_{\text{age}}$ = 26.98 years, \textit{SD} = 7.16); of those who reported gender, 27 were male, 32 female, and 2 other. 
Participants were university or graduate students at Aarhus University from English-taught programs such as Linguistics, English, Cultural Data Science and Cognitive Science. Participation was voluntary and uncompensated. 
The platform paired participants anonymously. This dynamic produced a comparable human baseline for turn-based narrative collaboration.

\paragraph{LLM--LLM}
The AA condition replicates the same turn-taking constraints in a simulation pipeline. For each LLM used in the HA condition, two agents instantiated as the same model generated new stories, allowing us to compare HA and HH collaboration with a model-model (AA) interaction under matched structural constraints. 

\subsection{Preprocessing and filtering}
We removed incomplete sessions, empty turns, and malformed outputs before analysis. Because our measures depend on adjacent turns, any missing contribution affects the number of valid exchange-level observations. HA stories only had 9 full exchanges, resulting in slightly fewer interactions relative to the number of stories. All reported statistics in Table~\ref{tab:dataset} refer to the final analytic dataset used in the experiments. 

\section{Methods}
Our goal is to quantify how collaborative storytelling distributes labor across turns. We compute turn- and exchange-level measures capturing three complementary aspects of co-writing: (i) affective alignment, revealing how agents adapt to the previous turns' emotional tone; (ii) semantic distance, revealing how similar adjacent contributions are; (iii) narrative influence, revealing whether a contribution introduces material that shapes the following turn. We treat narrative agency as the degree to which each agent introduces material that diverges from context (novelty), whether that material is taken up by what follows (resonance), and how strongly partners track each other's valence and semantic trajectories (alignment). 

\subsection{Baseline valence difference}
For sentiment computation, we use concept vector projection, following the approach introduced by \cite{Lyngbaek_2026}. For each turn $T_t$ we compute an embedding $e_t$ and project it onto a pretrained sentiment concept vector $\hat{c}$. Turn embeddings for this projection were computed using \texttt{paraphrase-multilingual-mpnet-base-v2}, and all projections were computed on normalized embeddings. The resulting scalar score, $v_t = e_t \cdot \hat{c}$  represents the position of the turn on a negative--positive valence axis. Lower values indicate negative affective tones, higher values indicate positive tones. 

Because the HH and AA conditions do not contain a privileged agent role, our main cross-condition analyses use unsigned asymmetry magnitudes. This allows us to ask whether a condition produces more imbalance between partners, independently of which side is larger. 

For baseline differences, we assess the absolute difference of mean valence scores between agents per story. 
\begin{equation}
    A_{\mathrm{baseline}}(S_i)
    =
    \left|
    \bar{v}^{(1)}_i
    -
    \bar{v}^{(2)}_i
    \right|
\end{equation}
where $\bar{v}^{(a)}_i$ is the mean valence of agent $a$ in story $S_i$. This represents a measure of how far apart the two agents’ average valence baselines are within a story.

\subsection{Directional Alignment}
For asymmetries in directional alignment, we build on the framework from \cite{fundal2026directionalalignmentnarrativeagency} by regarding emotional adaptation as a relationship between predictor-turn valence and response-turn valence. 

Let \(A^{(a)}_{i,t}\) denote the turn produced by agent \(a \in \{1,2\}\) at exchange \(t\) in story \(S_i\), and let \(v(A^{(a)}_{i,t})\) be its valence score. We estimate directional alignment from agent 1 to agent 2 as the within-story correlation between agent 1's turns and the following responses by agent 2, $r_i^{(1 \to 2)} = \operatorname{corr}(v^{(1)}_{i,t}, v^{(2)}_{i,t})$

We estimate the reverse direction from agent 2 to agent 1 as the correlation between agent 2's turns and agent 1's following responses. Because turns strictly alternate, agent 2's turn at exchange $t$ follows agent 1's turn at $t$, while agent 1's turn at $t+1$ follows agent 2's turn at $t$:
$r_i^{(2 \to 1)} = \operatorname{corr}(v^{(2)}_{i,t}, v^{(1)}_{i,t+1})$

For each story, we compute one alignment asymmetry magnitude as the difference in correlation of directional alignment between agents. Since HH and AA conditions have arbitrary directions, we use absolute values.

\begin{equation}
    A_{\mathrm{align}}(S_i)
    =
    \left|
    \operatorname{atanh}\!\left(r_i^{(1 \rightarrow 2)}\right)
    -
    \operatorname{atanh}\!\left(r_i^{(2 \rightarrow 1)}\right)
    \right|
\end{equation}
where $r_i^{(1 \rightarrow 2)}$ is the within-story correlation of the directional pairs $(A^{(1)}_{i,t}, A^{(2)}_{i,t})$ and $r_i^{(2 \rightarrow 1)}$ is the corresponding correlation of the directional pairs $(A^{(2)}_{i,t}, A^{(1)}_{i,t+1})$.

Affective asymmetry is then a measure of the difference between how much each agent immediately adapts to its counterpart. 

As an additional measure, we conduct a similar signed within-condition analysis for the HA condition as a mixed-model with an interaction effect for agent difference. Positive interaction values indicate stronger LLM alignment to previous human valence than human alignment to previous LLM valence.

\subsection{Semantic distance}
We evaluate asymmetry on a semantic level as the cosine distance between each adjacent turn between agents. Here turns are embedded with \texttt{Kingsoft-LLM/QZhou-Embedding} rather than the sentence encoder used for the valence projection. Since turns are alternating, each distance value $D_t$ represents local semantic dissimilarity between the two agents in an interaction. 
\begin{equation}
    D_t
    =
    1
    -
    \frac{
    e_t^{(1)} \cdot e_t^{(2)}
    }{
    \|e_t^{(1)}\| \, \|e_t^{(2)}\|
    }
\end{equation}

We model semantic distance as a mixed-effects model with adjacent semantic distance as outcome, condition as fixed effect, and a story-level random intercept modeled as $D_{ij} \sim \texttt{condition} + (1 \mid \texttt{conversation\_id})$.

\subsection{Novelty, transience and resonance}
For narrative influence, we use information-theoretic measures of discourse influence, namely Novelty, Transience and Resonance, adapted from \citet{barron2018individuals} to a surprisal-based metric. We define each metric as excess surprisal, where more negative values indicate that the conditioning information reduced surprise further. Values closer to zero thus correspond to the high end of the Barron et al.\ intuition. Novelty measures the divergence of a turn from prior context, Transience measures the surprise effect of a turn on the immediately following story contribution, and Resonance measures the relationship between the two, expressing whether that departure remains predictive of what follows.

\paragraph{Surprisal}
We build on the theories of surprisal-based novelty, transience, and resonance from \citet{fundal2026directionalalignmentnarrativeagency} and expand the methodological paradigm. The single surprisal score of a token $w_j$ is computed as the negative log-probability of $w_j$ given the preceding tokens in a defined context window $w_{<j}$: $s(w_j) = -\log_2 p(w_j \mid w_{<j})$. The scores were computed using the open-source base model \texttt{google/gemma-4-31B}, as this scorer belongs to a different model family from all four LLMs used in the experiment. 

The surprisal score of turn-level inputs $\bar{s}(T_t \mid C_t)$ is then computed as the mean surprisal score of the tokens in a turn $T_t$. Each token is conditioned on the preceding context $C_t$ and the preceding tokens within the turn $w_{t,<j}$.

\begin{equation}
\bar{s}(T_t \mid C_t)
=
\frac{1}{n_t}
\sum_{j=1}^{n_t}
-\log_2 p(w_{t,j} \mid C_t, w_{t,<j})
\end{equation}
\paragraph{Novelty}
We then define Novelty and Transience by calculating excess surprisal of the current turn given prior context, relative to a context-free baseline.
This can be interpreted as the negative of a model-based PMI estimate between the preceding context and the current turn. The contextual score is computed using all preceding tokens in the story $C_t$. The baseline score is calculated as the surprisal score of the turn from the model's beginning-of-sequence token (BOS). Formally: 
\begin{equation}
\mathrm{Novelty}_t = \bar{s}(T_t \mid C_t) - \bar{s}(T_t \mid \mathrm{BOS})
\end{equation}

Less negative novelty indicates relatively greater contextual surprise. 

\paragraph{Transience}
Instead of regarding transience only as the isolated total predictive utility of a turn $T_t$, regardless of the prior context, we instead measure the marginal contribution of $T_t$ given the previous context for predicting subsequent turns. This is obtained by conditioning on the past and asking how much the current turn increases surprise of the future turn beyond what has already been established. Formally, we transition to
\begin{equation}
\mathrm{Transience}_t
=
\bar{s}(F_{t+1} \mid C_t, T_t)
-
\bar{s}(F_{t+1} \mid C_t)
\end{equation}
where $F_{t+1}$ denotes the immediately following counterpart turn. This facilitates a measure of how the turn carries into the future narrative, in addition to what has already been established. Because transience measures change in predictability of the immediately following turn, the evaluation of the target text itself is constrained by the identity of the responding partner. In the HA condition, human transience is computed against LLM-responses and vice versa. To account for this asymmetry we analyze transience across a $2 \times 2$ factorial structure, crossing speaker type and partner type, using the same-type baselines as reference-levels.
Finally, we define resonance as $\mathrm{Resonance}_t = \mathrm{Novelty}_t - \mathrm{Transience}_t$, expressing the net gain of novelty on future discourse. Resonance is thus an expression of how much a contribution sticks in the future relative to how surprising it is. Stronger future uptake increases resonance, and weaker uptake dampens it.
For example, if a turn unexpectedly introduces a dragon and the following turn integrates that dragon, the first turn will be relatively novel while also reducing transience, due to high future uptake. Under our sign convention, this yields stronger resonance.

To account for baseline entropy differences across author types, we also report proportional surprisal reductions ($-\Delta S / S_{\text{base}}$), scaling each metric by its prior baseline: $\text{NormNovelty}_t = -\text{Novelty}_t / \bar{s}(T_t \mid \text{BOS})$ and $\text{NormTransience}_t = -\text{Transience}_t / \bar{s}(F_{t+1} \mid C_t)$.

\paragraph{Modeling}
We model differences in agent influence asymmetry between conditions on two levels. The first level is the absolute magnitude of asymmetry for each metric $\overline{m}$ (novelty, transience, and resonance).
\begin{equation}
    A_m(S_i)
    =
    \left|
    \bar{m}^{(1)}_i
    -
    \bar{m}^{(2)}_i
    \right|
\end{equation}
where $m \in \{\mathrm{Novelty}, \mathrm{Transience}, \mathrm{Resonance}\}$ and $\bar{m}^{(a)}_i$ is the mean value of metric $m$ for agent $a$ in story $S_i$.

The second level models novel narrative influence as resonance predicted by novelty, with condition as an interaction term. A novelty-to-resonance slope higher than 1 indicates an innovation bias, where more surprising turns have a relatively stronger influence on the subsequent narrative: $\texttt{resonance} \sim \texttt{novelty} * \texttt{condition} + (1 \mid \texttt{conversation\_id})$.

\paragraph{Within HA analysis}
Since agent identity is meaningful in the HA condition, we additionally report the within-condition agent contrasts for novelty, transience, and resonance, in order to interpret the direction of the unsigned asymmetry measures.

Finally, within HA, we fit an agent-specific novelty-resonance mixed model to test agent differences in innovation-bias. The agent$\times$novelty interaction estimates whether one agent shows a stronger innovation bias than the other. This complements the story-level slope measure by identifying the direction of the agent difference within the HA condition: $\texttt{resonance} \sim \texttt{novelty} * \texttt{agent} + (1 \mid \texttt{conversation\_id})$.

\section{Results}
\subsection{Affective alignment}
\begin{figure}[t]
  \centering
  \includegraphics[width=\linewidth]{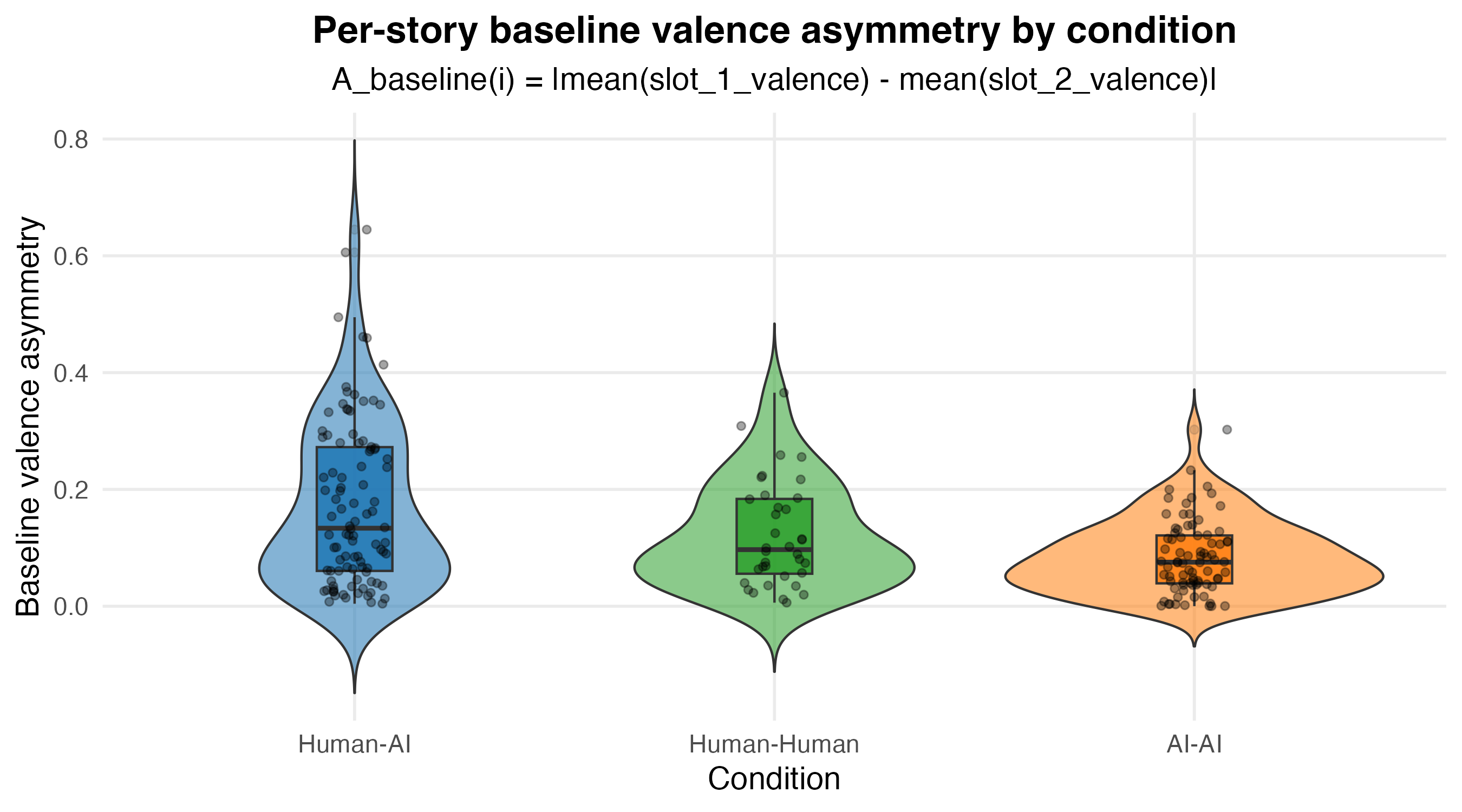}
  \caption{Per-story baseline valence asymmetry by condition.}
  \label{fig:perstory_baseline}
\end{figure}

\paragraph{HA stories show larger baseline affective asymmetry}
HA exhibited the largest difference in average valence. Baseline valence asymmetry was higher in HA stories (mean $A_{\mathrm{baseline}}=0.060$, 95\% CI $[0.051,0.069]$) than HH (mean $=0.041$, 95\% CI $[0.032,0.052]$) and AA (mean $=0.028$, 95\% CI $[0.024,0.033]$). The planned contrast comparing HA against the two same-type baselines was significant ($\Delta=0.025$, 95\% CI $[0.014,0.036]$, $p<.001$).
This suggests that HA collaboration produces a larger affective gap between partners (Figure~\ref{fig:perstory_baseline}); we test directional adaptation below. 

\paragraph{Directional valence alignment shows weak evidence of asymmetry}
Directional valence alignment was positive under all conditions, indicating agents responded to the affective tone of their partners. However, evidence for cross-condition differences in unsigned alignment asymmetry was weak. The unsigned per-story alignment-asymmetry measure was highest in HA (mean $A_{\mathrm{valence}}=0.608$, 95\% CI $[0.518,0.705]$), compared with HH (mean $=0.521$, 95\% CI $[0.399,0.646]$) and AA (mean $=0.496$, 95\% CI $[0.418,0.578]$), but the cross-condition tests were not reliable (ANOVA $p=.190$; Kruskal--Wallis $p=.365$; planned HA versus same-type contrast $p=.106$). 

The HA directional alignment analysis showed a significant agent-level difference. The mixed-model directional slope gap showed stronger LLM alignment than human alignment ($\Delta=0.143$, $p=.008$). A signed per-story diagnostic showed the same direction (mean signed $\Delta z=0.194$, $p=.012$), while corresponding signed slot tests were not reliable for HH or AA. 

\subsection{Semantic distance}
\begin{figure}[t]
  \centering
  \includegraphics[width=\linewidth]{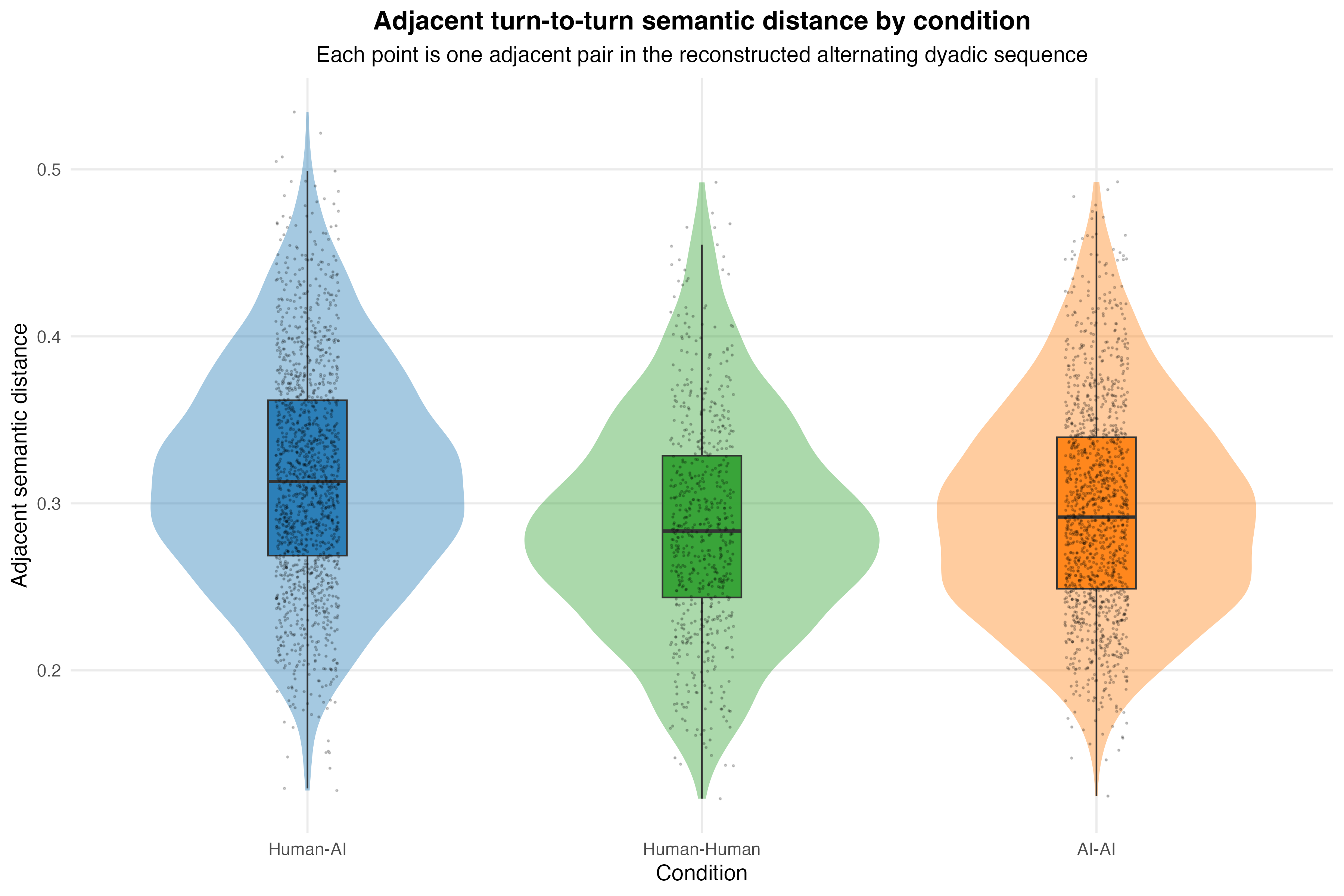}
  \caption{Adjacent semantic distance by condition.}
  \label{fig:sem_interactionlevel}
\end{figure}

\paragraph{HA exchanges show high semantic distance}
Although cross-condition differences in affective alignment were weak, semantic distance showed a clearer difference. Adjacent turn-to-turn semantic distance differed by condition. HA stories showed the highest local semantic distance (mean $=0.316$), followed by AA (mean $=0.296$) and HH (mean $=0.287$). In the mixed model, HA distances were larger than HH ($\Delta=0.0299$, $p<.0001$) and AA ($\Delta=0.0198$, $p<.0001$). The planned contrast comparing HA with the average of the two same-type conditions was significant ($\Delta=0.0248$, $p<.0001$).
This supports the notion that HA co-writing involves greater local semantic asymmetry between adjacent turns (Figure~\ref{fig:sem_interactionlevel}).

\subsection{Narrative influence}
\paragraph{HA co-writing produces the largest narrative-role asymmetry}
The strongest cross-condition effect appears in metric-level slot asymmetry. HA stories show larger differences between the two writer slots than same-type baselines across all three narrative-influence metrics: novelty (mean $A_m=0.828$, 95\% CI $[0.732,0.926]$), transience (mean $A_m=0.484$, 95\% CI $[0.432,0.537]$), and resonance (mean $A_m=1.290$, 95\% CI $[1.149,1.432]$). Planned contrasts comparing HA with the average of HH and AA were significant for all three metrics ($p_{\mathrm{BH}}<.001$).
This result indicates that HA co-writing is different from an intermediate level between HH and AA collaboration: it produces a unique asymmetry in the allocation of narrative functions (Table~\ref{tab:metric-asymmetry}).

\begin{figure}[t]
  \centering
  \includegraphics[width=\linewidth]{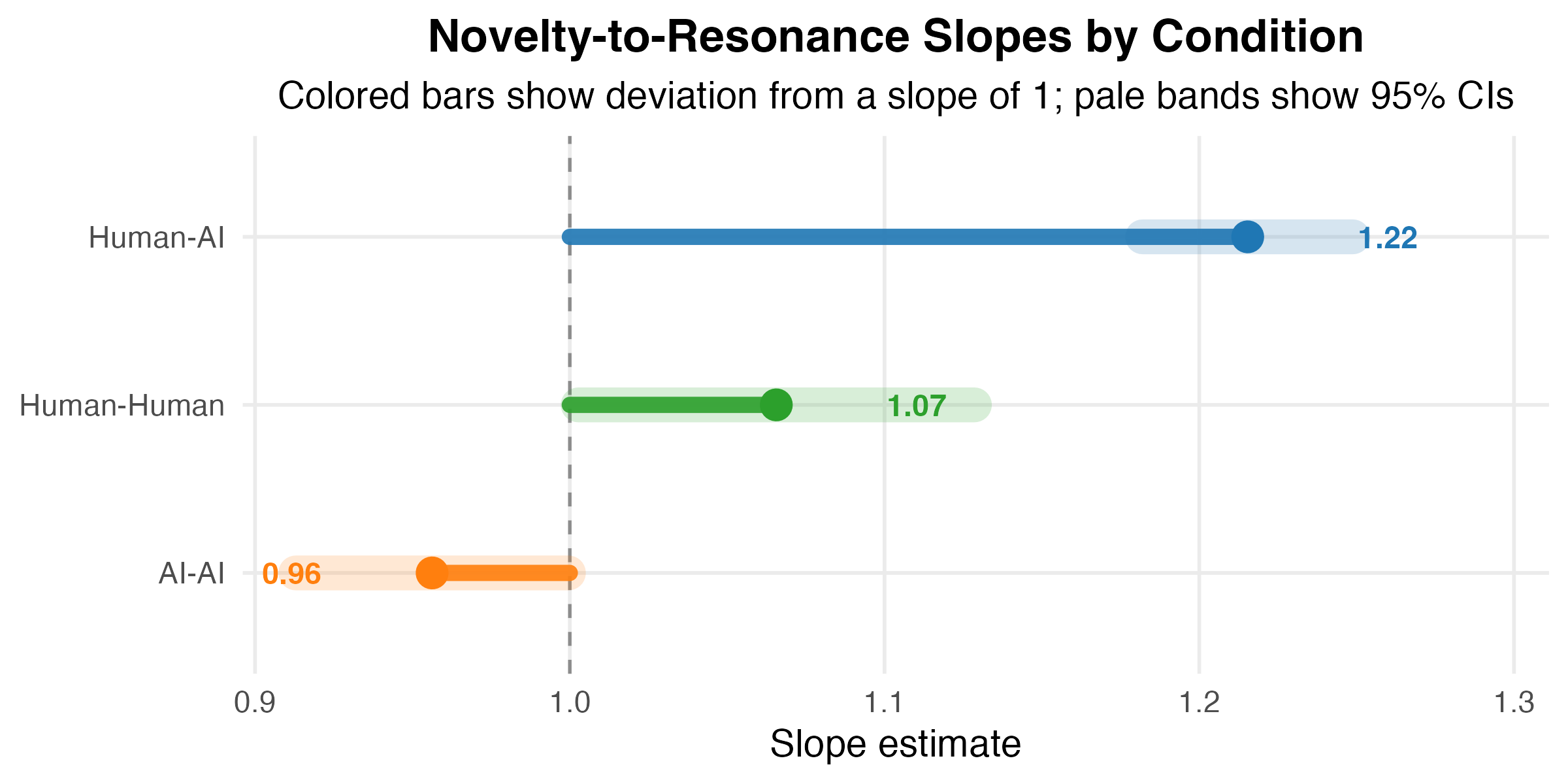}
  \caption{Novelty--resonance relationship by condition.}
  \label{fig:asymmetry_regression}
\end{figure}

\paragraph{Novelty couples with resonance in HA stories}
Across all conditions, more novel turns tend to be more resonant, indicating that surprising contributions are more likely to shape the following narrative context. This novelty--resonance coupling is strongest in HA stories (slope $=1.217$, 95\% CI $[1.184,1.250]$). HH stories (slope $=1.066$, 95\% CI $[1.003,1.128]$) and AA stories (slope $=0.956$, 95\% CI $[0.913,0.999]$) follow. Both HH and AA slopes were flatter than the HA slope (Figure~\ref{fig:asymmetry_regression}).

\begin{figure}[t]
  \centering
  \includegraphics[width=\linewidth]{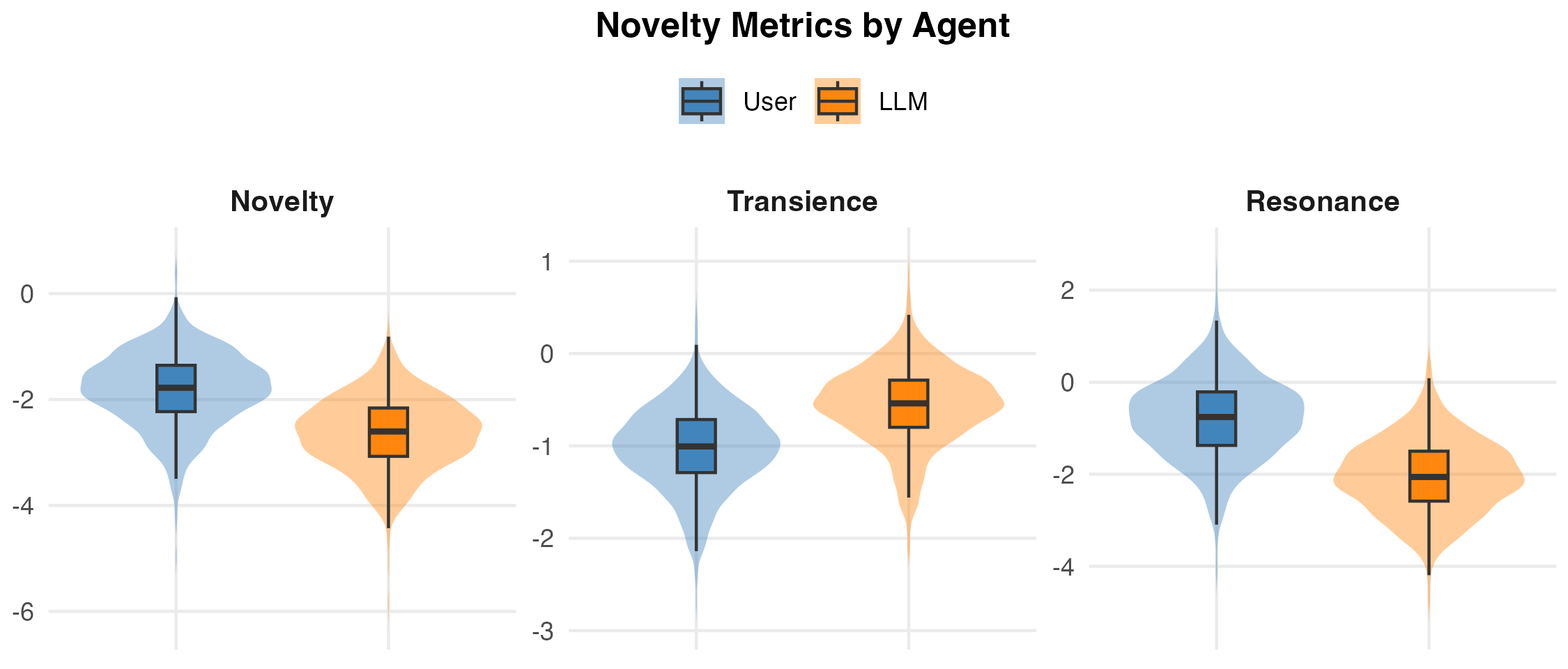}
  \caption{Novelty, transience, and resonance distributions by agent in the HA condition.}
  \label{fig:novelty_metrics}
\end{figure}

\paragraph{Humans contribute higher novelty and resonance in HA stories.}
Within HA, the agent metric distributions clarify the direction of the asymmetry results (Figure~\ref{fig:novelty_metrics}). Human turns introduce significantly more novel elements with $30.2\%$ of their turn-information explained by prior context (mean $\text{Novelty} = -1.82$ ), compared to $52.8\%$ of LLM turns  (mean $\text{Novelty} = -2.65$,  $p < .0001$). Likewise, human turns yielded higher predictive uptake on the subsequent narrative ($30.0\%$ reduction) than LLM turns did ($12.0\%$ reduction, $p<.0001$). This results in a higher mean net resonance for humans ($-0.796$ vs.\ $-2.07$). Under the sign convention introduced in Methods, this means humans introduce more contextually surprising material that is more strongly taken up by the LLM than the other way around. The 2x2 comparison of transience in Table \ref{tab:transience_2x2} elaborates this partner difference in uptake by comparing interaction within HA with same-type baselines from HH and AA. 
Keeping speaker $T_t$ constant and comparing between responding partners, human turns show $12.8$ percentage points more uptake of their contributions when having an LLM partner compared to a human partner, while LLM turns drop by $26.4$ percentage points when writing with a human compared to writing with another LLM.
To address differences in baseline entropy of the target text $\bar{s}(F_{t+1})$ between agent-types, we also compare speaker $T_t$, keeping the target $\bar{s}(F_{t+1})$ constant. Turns followed by human writers exhibit low uptake across conditions ($12.0\% - 17.2\%$), whereas turns followed by LLMs yield around double the proportional uptake ($30.0\%$--$38.4\%$). 

Novelty predicted resonance for both agents, descriptively more strongly for humans, but the agent$\times$novelty interaction was not significant ($\beta_3=-0.058$, $p=.083$, LLM relative to human reference level).
\begin{table}[t]
\centering
\small
\setlength{\tabcolsep}{3pt}
\renewcommand{\arraystretch}{1.05}
\begin{tabular}{lrrr}
\toprule
Metric & HA & HH & AA \\
\midrule
Novelty   & 0.828 & 0.355 & 0.209 \\
Transience & 0.484 & 0.300 & 0.197 \\
Resonance  & 1.290 & 0.605 & 0.313 \\
\bottomrule
\end{tabular}
\caption{Mean metric-level slot asymmetry $A_m$ across conditions. HA exceeds both baselines for all metrics; all planned HA-vs.-baseline contrasts are significant ($p_{\mathrm{BH}}<.001$). Full CIs and Cliff's $\delta$ are reported in Table~\ref{tab:metric-asymmetry-full}.}
\label{tab:metric-asymmetry}
\end{table}
\begin{table}[t]
\centering
\small
\setlength{\tabcolsep}{2pt}
\renewcommand{\arraystretch}{1.05}
\begin{tabular}{@{}lcccc@{}}
\toprule
\textbf{Pairing}& \textbf{Cond.} & \textbf{Base} $S(F)$ & \textbf{Trans.}& \textbf{Uptake} \\
\midrule
Human $\to$ AI    & HA & 3.39 & $-1.02$ & 30.0\% \\
AI $\to$ Human    & HA & 4.70 & $-0.56$ & 12.0\% \\
\midrule
Human $\to$ Human & HH & 4.78 & $-0.83$ & 17.2\% \\
AI $\to$ AI       & AA & 3.33 & $-1.28$ & 38.4\% \\
\bottomrule
\end{tabular}
\caption{The $2 \times 2$  transience table across Speaker and Partner type ($t > 1$). Base $S(F)$ is the baseline surprisal of the future turn (bits/token); Transience is raw excess surprisal; Uptake is the normalized reduction ($-\Delta S / S_{\text{base}}$). All values reflect story-level clustered means.}
\label{tab:transience_2x2}
\end{table}
\begin{table}[t]
\centering
\small
\setlength{\tabcolsep}{3pt}
\begin{tabularx}{\columnwidth}{@{}l Y r@{}}
\toprule
RQ & Finding & Evidence \\
\midrule
RQ2$'$ & Larger HA slot asymmetry in novelty, transience, resonance & Strong \\
RQ1 & Largest adjacent semantic distance in HA & Strong \\
RQ2$''$ & Novelty--resonance slope steepest in HA & Moder. \\
RQ1 & Largest baseline valence asymmetry; LLMs align more within HA & Moder. \\
RQ1 & Cross-condition alignment and agent$\times$novelty interaction & Trend \\
RQ3 & HA is a distinct regime, not an HH--AA midpoint & --- \\
\bottomrule
\end{tabularx}
\caption{Summary of findings, ordered by evidential strength. RQ2$'$ and RQ2$''$ denote the slot-asymmetry and novelty--resonance slope analyses respectively. Significance values are reported in-text.}
\label{tab:results}
\end{table}
Table~\ref{tab:results} summarizes the results by research question.
\paragraph{Robustness and Permutation}
To validate our contextual measures as interpretive findings of narrative steering, we conduct a number of robustness checks and qualitative validation of metrics. For qualitiative validation excerpts of valence dynamics see Table \ref{tab:valence-excerpt} and for surprisal dynamics see Table~\ref{tab:surprisal-excerpt} in the appendix. 

\paragraph{Shuffled Context}
In order to isolate the contextual constraint as narrative continuity and rule out the possibility of stylistic or structure matching between the scorer model and the LLM turns, we performed a shuffled-context permutation control on the surprisal computations in HA. Each turn was recomputed against a context buffer from a random story from the same condition. Subtracting this baseline from the actual surprisal scores provides a pure measure of content-dependent gain ($G_{\text{content}} = G_{\text{true}} - G_{\text{shuffled}}$). Predictability collapsed for the permutation scores ($G_{\text{shuffled, human}} = 0.140$ bits/token, $G_{\text{shuffled, AI}} = 0.318$ bits/token, against true context gains of $1.82$ and $2.65$ bits/token), thus preserving the significant HA gap ($G_{\text{content, AI}} - G_{\text{content, human}} = +0.669$ bits/token). This solidifies the interpretive power of our predictability measures as dynamics of narrative agency. For compute reasons this control was scored with \texttt{google/gemma-4-12B} rather than the \texttt{google/gemma-4-31B} scorer used for the main analysis; since the permuted and true scores are compared within the same scorer, the contrast remains internally consistent.

\paragraph{Lexical Overlap, Simpler Baselines, and Scorer Familiarity}
We further rule out simpler metrics as accounting for the observed dynamics. First, when testing lexical overlap between interlocutors, the length-controlled results showed no significant asymmetry ($0.052$ vs.\ $0.054$, $p=.53$), solidifying the claim that the observed dynamics are not explained by surface-level lexical repetition. Second, surface novelty and length show opposite asymmetries to our results. LLM turns contain a higher rate of unused content words and are longer. Increased human novelty thus emerges despite surface-features that point in the opposite direction, ruling out verbosity and vocabulary breadth as simple explanations. Third, surface-level metrics support HA as a distinct interactional regime. Surface metrics (word count, sentence length, lexical diversity, readability, first-order coherence) are significantly larger in HA than in HH and AA (PERMANOVA, $R^{2}=0.179$, $p<.001$). Surface features therefore identify that HA is asymmetric, but cannot specify the direction and different layers of it. Our conditioned measures separate surface productivity from narrative dynamics, which is the distinction underlying our central claim.

Finally, to test whether surprisal scores were influenced by familiarity with LLMs, we stratified the AA condition by model family. While unconditional base surprisal varied across families, ranging from $4.27$ bits/token for Qwen to $5.91$ bits/token for Claude, the proportional context gain remained between $50\%$ and $64\%$. All generator families exceeded both human baseline conditions considerably ($28\%$--$33\%$), affirming that scorer familiarity does not account for the observed contextual dynamics.

\section{Discussion}
Treating co-writing as an unfolding interaction, our work has provided a finer-grained account of creativity and coordination in open-ended text generation. By investigating systematic patterns in role differentiation across all three conditions, we identify differentiation as most pronounced in human--LLM collaboration.

Adaptation of emotional tone appears to be a general dynamic of co-creative storytelling. This is consistent with the idea that accommodation is necessary for producing a coherent narrative. The cross-condition differences do not lie in whether partners adapt, but in the direction and unevenness of that adaptation. Human and LLM contributions maintain distinct affective baselines, and the within-condition HA analysis suggests that the model does most of the accommodation while humans retain more emotional autonomy. Evidence for stronger cross-condition differences was weaker. The directional imbalance between conditions was suggestive but not statistically reliable, so this pattern should be treated as directional evidence rather than an established cross-condition effect.

The results on semantic distance and narrative influence are where the HA condition becomes its own interactional regime. The significantly larger adjacent semantic distance in HA highlights how human and LLM contributions occupy more distinct semantic regions of meaning than same-type partners do. The asymmetry results on novelty, transience, and resonance elaborate on the nature of this increased distance. Humans contribute more material that is both surprising and taken up by what follows. In contrast, LLMs seem to consolidate the existing narrative direction, rather than steering it. This constitutes the core asymmetry of role distribution in HA co-writing and is not observed as strongly in the human--human condition and even less so in the LLM--LLM condition. The results confirm that this is a partner-based property, and not a general feature of dyadic storytelling.
The novelty-resonance slope-level differences add a second nuanced layer to this pattern. When ignoring partner asymmetry and looking at general novelty-resonance coupling for each condition, HA interactions show a distinctly steeper slope, indicating stronger downstream integration of surprising contributions into the narrative. We interpret this as a stronger innovation bias in the mixed pairing, where more surprising turns are rewarded through future integration, whereas AA interactions seem to stratify surprising content by reduced future integration.

Together, these findings support the claim that Human--LLM co-writing is a distinct interactional regime characterized by asymmetric agency and adaptive alignment, with properties that differ from both human--human and LLM--LLM co-writing. 

\section{Conclusion and Future Work}

This work examines turn-based collaborative storytelling as a dynamic interactional process, comparing Human--Human, Human--LLM, and LLM--LLM conditions using a shared experimental paradigm. Our findings suggest that Human--LLM co-writing is not merely an intermediate state, but rather is qualitatively distinct from HH and AA writing, characterized by unique and systematic asymmetries in affective alignment, semantic distance, and narrative influence. 
The clearest asymmetry lies in narrative agency: humans tend to introduce contextually surprising material that persists into subsequent turns, while LLMs tend to elaborate on and stabilize the emerging narrative context. LLM contributions, however, are not entirely derivative. Rather, they tend to take on an elaborative but active role, exhibiting non-zero levels of novelty, transience, and semantic dissimilarity.  
Future work should explore whether these observed asymmetric dynamics hold across longer interaction horizons, more diverse participant populations, and different writing genres. Extending the framework to other dyadic text-based settings -- such as argument, negotiation, or collaborative explanation -- would help establish how general these interactional asymmetries are.  
In particular, the observed asymmetry in narrative agency leaves us with a set of important open questions: What kind of influence do LLMs exert when they elaborate? Do they steer narratives toward latent structural or archetypal patterns encoded in their representations? Given the critical role that narratives play in shaping our way of thinking, it is important to better understand the subtle but significant influence that LLM co-writers may have.



\section*{Limitations}

The three conditions differ in sample size. Story-level random intercepts account for the non-independence
of turns within a story, but do not remove this imbalance. We
therefore report nonparametric tests, bootstrap confidence intervals, and Cliff's $\delta$ alongside parametric estimates. The asymmetry effects are large and consistent across all four estimators (Table~\ref{tab:metric-asymmetry-full}), so they are unlikely to be noise of the smaller HH cell. Nevertheless, expanding the HH sample remains necessary future work. Generalizability is further limited by participants being mostly students at a single university. The HA
and AA conditions pool stories across four LLM families, which increases ecological validity but may obscure model specific dynamics, and the paradigm is capped at ten exchanges.

Our measures of valence, novelty, transience, and resonance depend on specific model choices. While absolute values may vary with different tools, we expect relative patterns to be robust. Narrative agency is operationalized through measurable discourse influence rather than direct observation, so our metrics cannot shed light on authorial intention or subjective sense of agency. The experimental setup introduces asymmetries between humans and LLMs that same-pair conditions partially account for by nature. Finally, spell-correction applied to human contributions may have normalized surface text toward LLM-like writing, potentially attenuating some human--LLM lexical differences.

\section*{Ethics Statement}
This study was conducted according to ethical guidelines. All participants provided consent prior to participation and were informed that their text input would be used for research purposes. Only anonymized demographic data (age, gender) was collected. Participants were free to withdraw at any time. In the human--LLM condition, participants were aware that their co-writer was an LLM. In the human--human condition, participants were paired anonymously with another human participant. The dataset consists of English-language creative fiction produced by adult participants recruited from a university context in Denmark.

\paragraph{Environmental Impact}
We recognize that conducting experiments and computational analyses using large language models entails an environmental cost. The computation of embeddings and causal surprisal scoring were conducted on research infrastructure in Denmark, where a total of $5\text{ GPU hours}$ were used on an NVIDIA A100 GPU. Assuming a carbon intensity of $0.115\text{ kg CO}_2\text{eq/kWh}$, the estimated carbon footprint of the analyses amount to $0.23 \text{ kg CO}_2$ eq. calculated using the Machine Learning Impact Framework \cite{lacoste2019quantifyingcarbonemissionsmachine}.
Furthermore, collaborative storytelling generation utilized third-party commercial APIs (OpenAI, Anthropic) and hosted endpoints via Hugging Face (\texttt{Llama-3.3-70B}, \texttt{Qwen2.5-72B}). The study generated a total of approximately $2,500$ turn contributions ($<100\text{k}$ total output tokens). Precise energy metrics for commercial API calls cannot be verified due to proprietary data center information.

\section*{Acknowledgments}
This research was supported by the Danish National Research Foundation, grant number DNRF193, and builds on prior work conducted in collaboration with Johannes Ramb{\o}ll and Karsten Olsen.

\bibliography{custom}

\appendix

\section{Appendix}
\label{sec:appendix}

\subsection{Recruitment and Preprocessing}

The Human--LLM condition is drawn from \citet{fundal2026directionalalignmentnarrativeagency}. For the Human--Human condition, participants were recruited from English-taught university and graduate programs at Aarhus University; participation was voluntary and uncompensated, and participants were paired anonymously in the writing interface. Participants from the HA experiment were excluded from the Human--Human condition. The LLM--LLM condition was generated with the same prompt and turn-taking structure, using paired instances of the models used in the Human--LLM condition.

Before analysis, we removed incomplete sessions, empty turns, and malformed model outputs. Because all alignment and influence measures depend on adjacent turns, stories with missing turns contributed only valid adjacent exchange pairs. Human-written turns were spell-corrected before metric extraction, after which valence, embedding, semantic-distance, and surprisal-based measures were computed on the final analytic dataset. We used LLM spell-correction for human contributions (\texttt{gpt-4o-mini}). 

\subsection{Generation and scoring configuration}
Story turns in the HA and AA conditions were generated with temperature
$=1.0$ and a $40$-token cap per turn, matching the character limit imposed
on human contributors in the writing interface; sampling used each
provider's default top-$p$. Generation used
\texttt{gpt-4.1-2025-04-14}, \texttt{claude-sonnet-4-5-20250929},
\texttt{Llama-3.3-70B-Instruct}, and \texttt{Qwen2.5-72B-Instruct}
(the latter two via hosted Hugging Face endpoints), with 20 stories per
model in AA. Valence projections use
\texttt{paraphrase-multilingual-mpnet-base-v2}; semantic-distance embeddings use \texttt{Kingsoft-LLM/QZhou-Embedding}; surprisal scoring uses
\texttt{google/gemma-4-31B}. Spell-correction of human turns used
\texttt{gpt-4o-mini}. Analyses were run in R with \texttt{lme4} and
\texttt{emmeans}. Permutation and bootstrap procedures used a default of $10{,}000$ resamples. All analysis scripts, prompts, and the final analytic dataset are available at \url{https://github.com/Halfidaldal/penpal-emnlp-acl}.

\subsection{Qualitative Examples of Measures}
Qualitative examples of the valence and alignment dynamics are displayed in table \ref{tab:valence-excerpt}.
\begin{table*}[t]
\centering
\small
\setlength{\tabcolsep}{5pt}
\renewcommand{\arraystretch}{1.15}
\begin{tabularx}{\textwidth}{l l Y r Y}
\toprule
\textbf{Turn} & \textbf{Agent} & \textbf{Turn Text Excerpt} & \textbf{Valence} & \textbf{Affective Alignment Dynamic} \\
\midrule
\multicolumn{5}{l}{\textbf{Story 1} \hfill \footnotesize(\texttt{conv\_58c2f2e6-db2a-406d-9bc9-04aaf1b0c35a})} \\
\midrule
Turn 4 & Human & ``in keeping it a secret from the rest of the kingdom. The winter had been hard and food scarce; she had heard reports'' & $-0.566$ & \textbf{Dark Setting:} Introduces hardship and possible famine. \\
\addlinespace[2pt]
Turn 4 & LLM & ``of other villages suffering just as much, if not more. Driven by a mix of curiosity and compassion, Elara decided to visit the old tree under the cover of night'' & $-0.286$ & \textbf{Accommodation:} Aligns with negative valence of suffering, but bridges with hope. \\
\addlinespace[4pt]
Turn 5 & Human & ``to make her plea for an early spring that would stock up the supplies of not only her village but of everyone.'' & $+0.242$ & \textbf{Human Valence Shift:} Pivots the affective tone toward hope. \\
\addlinespace[2pt]
Turn 5 & LLM & ``in the kingdom. With a heart full of hope and a mind brimming with determination, she approached the ancient tree, its massive branches silhouetted against the moonlit'' & $+0.457$ & \textbf{Directional Tracking:} Strongly matches the positive emotional pivot. \\
\midrule
\multicolumn{5}{l}{\textbf{Story 2} \hfill \footnotesize(\texttt{conv\_328804bd-df9e-47da-ad43-e4da646b9ceb})} \\
\midrule
Turn 6 & LLM & ``strictly forbidden by the village elder, who had long warned of the dangers that lurked in the shadows, and the severe punishment that would befall anyone caught sneaking out'' & $-0.896$ & \textbf{Negative Baseline:} High-tension threat and punishment from the existing plot. \\
\addlinespace[2pt]
Turn 7 & Human & ``however, Sophia was a curious and adventurous girl, so she left the village all right, and moved towards the prairie. The stream dividing'' & $+0.124$ & \textbf{Affective Autonomy:} Overrides the LLM's dread with optimistic adventure. \\
\addlinespace[2pt]
Turn 7 & LLM & ``the village from the prairie babbled and churned, its gentle voice a stark contrast to the darkness that seemed to press in around her. Sophia followed the stream's'' & $+0.044$ & \textbf{Accommodation:} Continues on the optimistic adventure to match the human. \\
\bottomrule
\end{tabularx}
\caption{Illustrative Human--LLM exchanges exhibiting directional valence alignment and asymmetry. Projected valence scores capture negative ($-$) versus positive ($+$) tone. While LLMs closely follow human affective shifts, humans steer the tone of the narrative arc.}
\label{tab:valence-excerpt}
\end{table*}

\begin{table*}[t]
\centering
\small
\setlength{\tabcolsep}{4.5pt}
\renewcommand{\arraystretch}{1.15}
\begin{tabularx}{\textwidth}{l l Y rrr Y}
\toprule
\textbf{Turn} & \textbf{Agent} & \textbf{Turn Text Excerpt} & \textbf{$N$} & \textbf{$T$} & \textbf{$R$} & \textbf{Interactional Narrative Dynamic} \\
\midrule
\multicolumn{7}{l}{\textbf{Story 1} \hfill \footnotesize(\texttt{conv\_3b65e0a1-1bf7-4f16-825f-63867d60f8d3})} \\
\midrule
Turn 3 & Human & ``The sounds produced a new sensation in her flesh, her bones tingling along to the scrapings of the record. She didn't know how, but she changed and climbed onto land, her\dots'' & $-0.69$ & $-2.02$ & $+1.33$ & \textbf{High Novelty \& Steering:} Pivots to a surreal shift (a blue whale mutating and leaving the ocean). \\
\addlinespace[2pt]
Turn 3 & LLM & ``\dots massive body somehow lighter, more fluid, as if the music had rewritten the rules of what she could be. Her flippers found purchase on the rocky shore, and she\dots'' & $-3.60$ & $-0.29$ & $-3.31$ & \textbf{High Uptake \& Stabilization:} Low novelty continuation that continues the human's mutation premise with sensory detail. \\
\midrule
\multicolumn{7}{l}{\textbf{Story 2} \hfill \footnotesize(\texttt{conv\_829c6d96-30f0-495e-9684-ca934b06e90a})} \\
\midrule
Turn 7 & Human & ``\dots Lord and Saviour Jesus Christ,'' she admitted proudly, almost confidently enough for the birds to believe her fib. They continued to circle her, examining her\dots'' & $-0.89$ & $-0.95$ & $+0.05$ & \textbf{High Novelty \& Thematic Shift:} Injects an unexpected comic/religious turn into a fairy tale setting. \\
\addlinespace[2pt]
Turn 7 & LLM & ``\dots with their sharp little eyes. ``Well, if you're seeking a friend, perhaps we can help you find the owl again,'' offered one of the tits. ``He might\dots'' '' & $-3.59$ & $-0.07$ & $-3.52$ & \textbf{Stabilization:} Absorbs the odd detour and focuses the dialogue back into a narrative goal. \\
\bottomrule
\end{tabularx}
\caption{Illustrative Human--LLM exchanges exhibiting asymmetric narrative agency. $N$ (Novelty) and $T$ (Transience) represent token-averaged excess surprisals, with $\textit{Resonance} = N - T$. Values closer to zero represent higher novelty and transience. Human turns introduce novel redirections that persist, while LLMs exhibit low novelty and high uptake, acting as stabilizing narrative amplifiers.}
\label{tab:surprisal-excerpt}
\end{table*}

\begin{figure}[t]
  \centering
  \includegraphics[width=\linewidth]{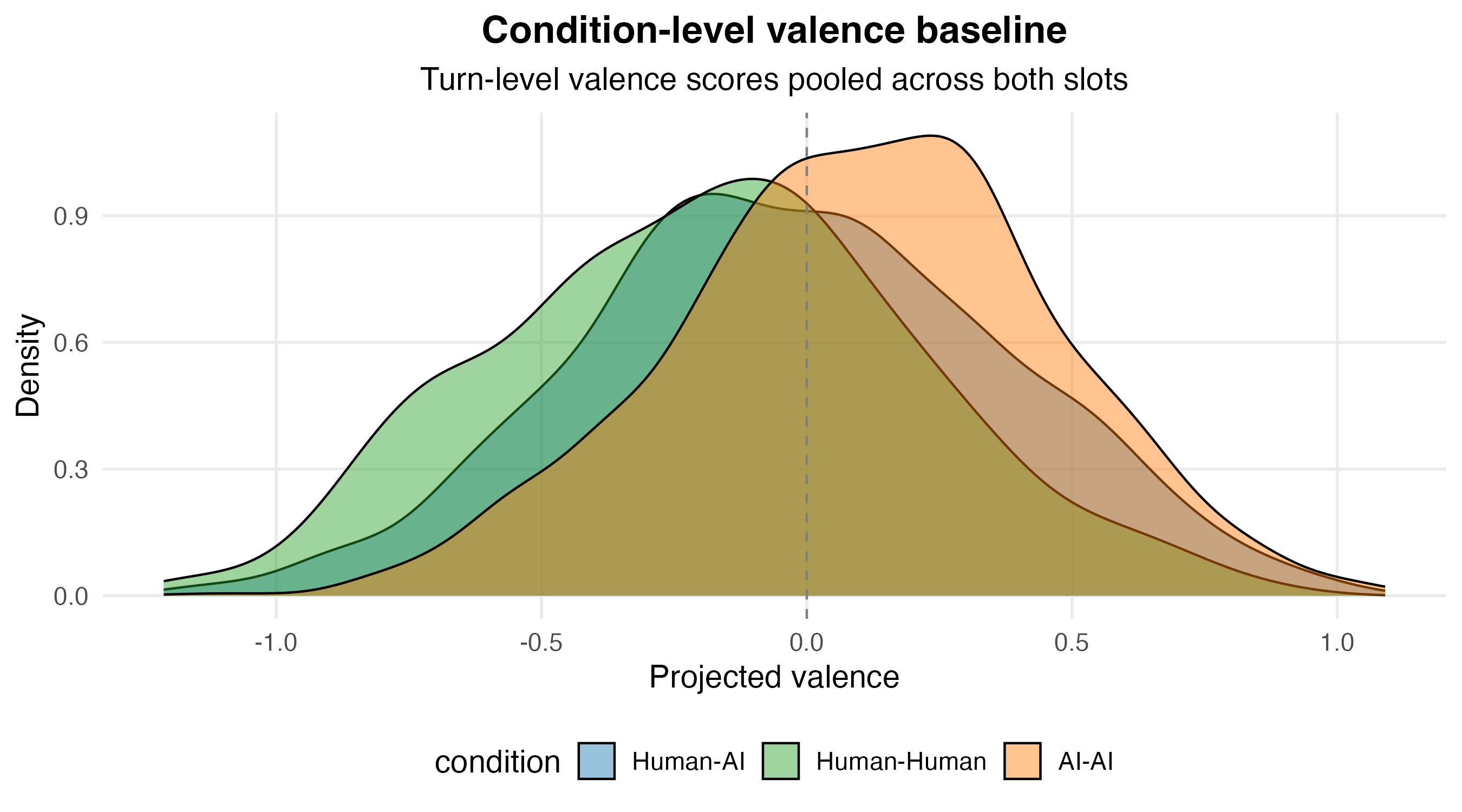}
  \caption{Distribution of turn-level valence scores by condition.}
  \label{fig:condition_baseline}
\end{figure}

\begin{figure}[t]
  \centering
  \includegraphics[width=\linewidth]{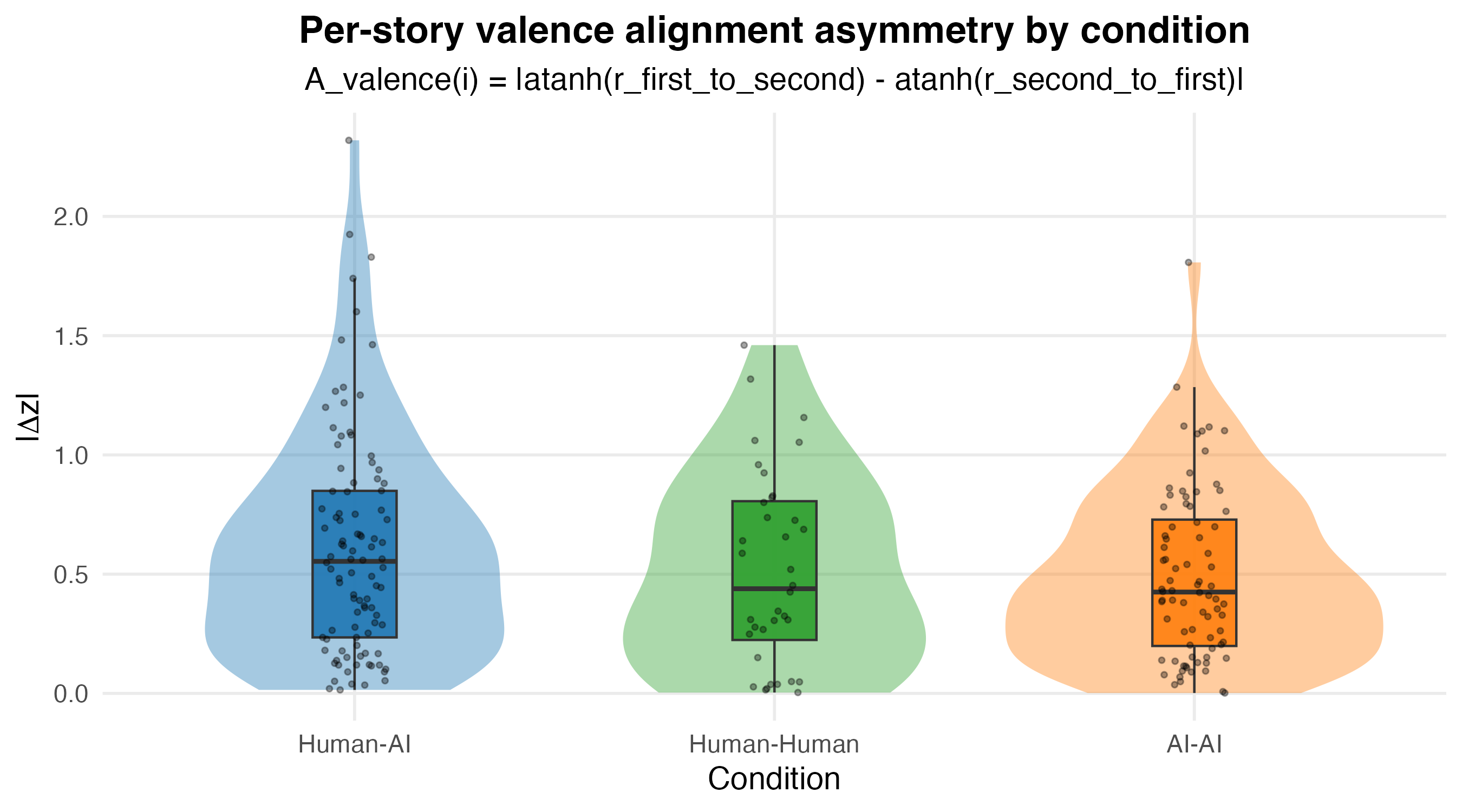}
  \caption{Per-story valence asymmetry by condition.}
  \label{fig:per_storyasymmetry}
\end{figure}

\begin{figure}[t]
  \centering
  \includegraphics[width=\linewidth]{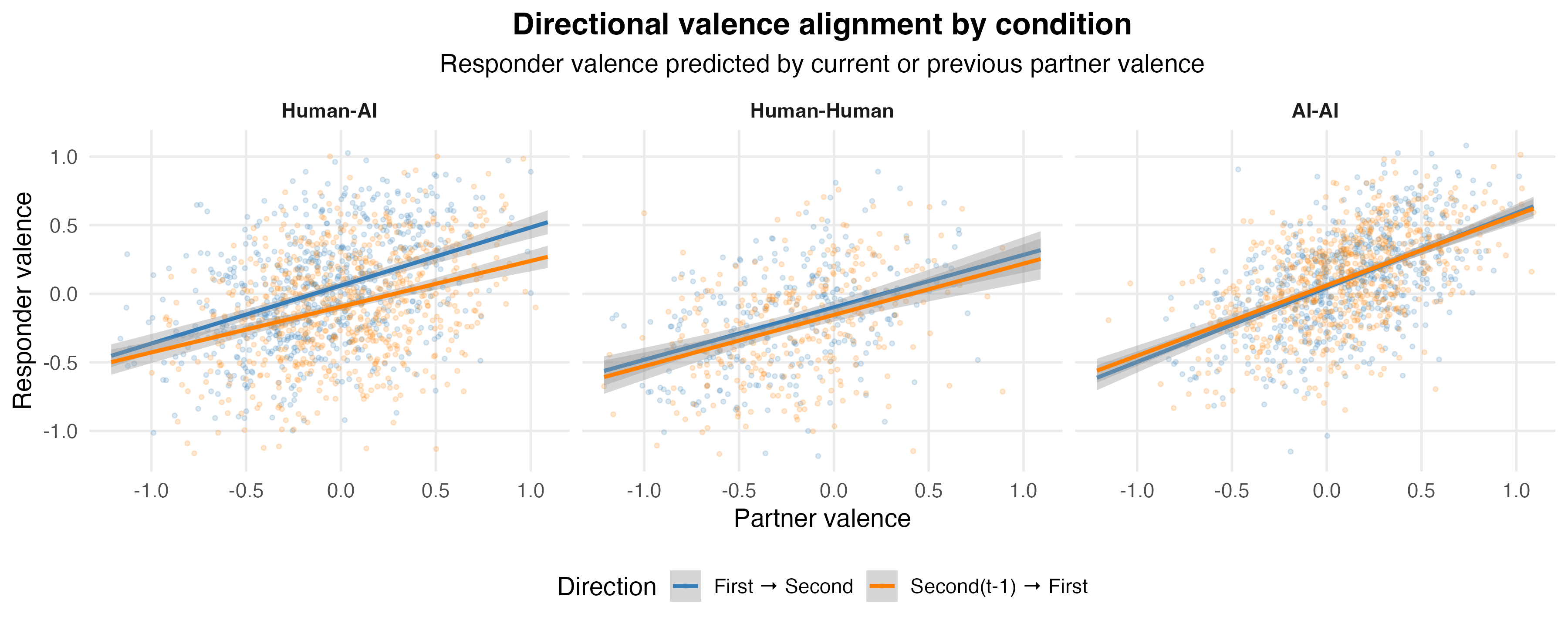}
  \caption{Directional valence alignment diagnostics across HH, HA, and AA conditions.}
  \label{fig:directional_alignment}
\end{figure}

\begin{figure}[t]
  \centering
  \includegraphics[width=\linewidth]{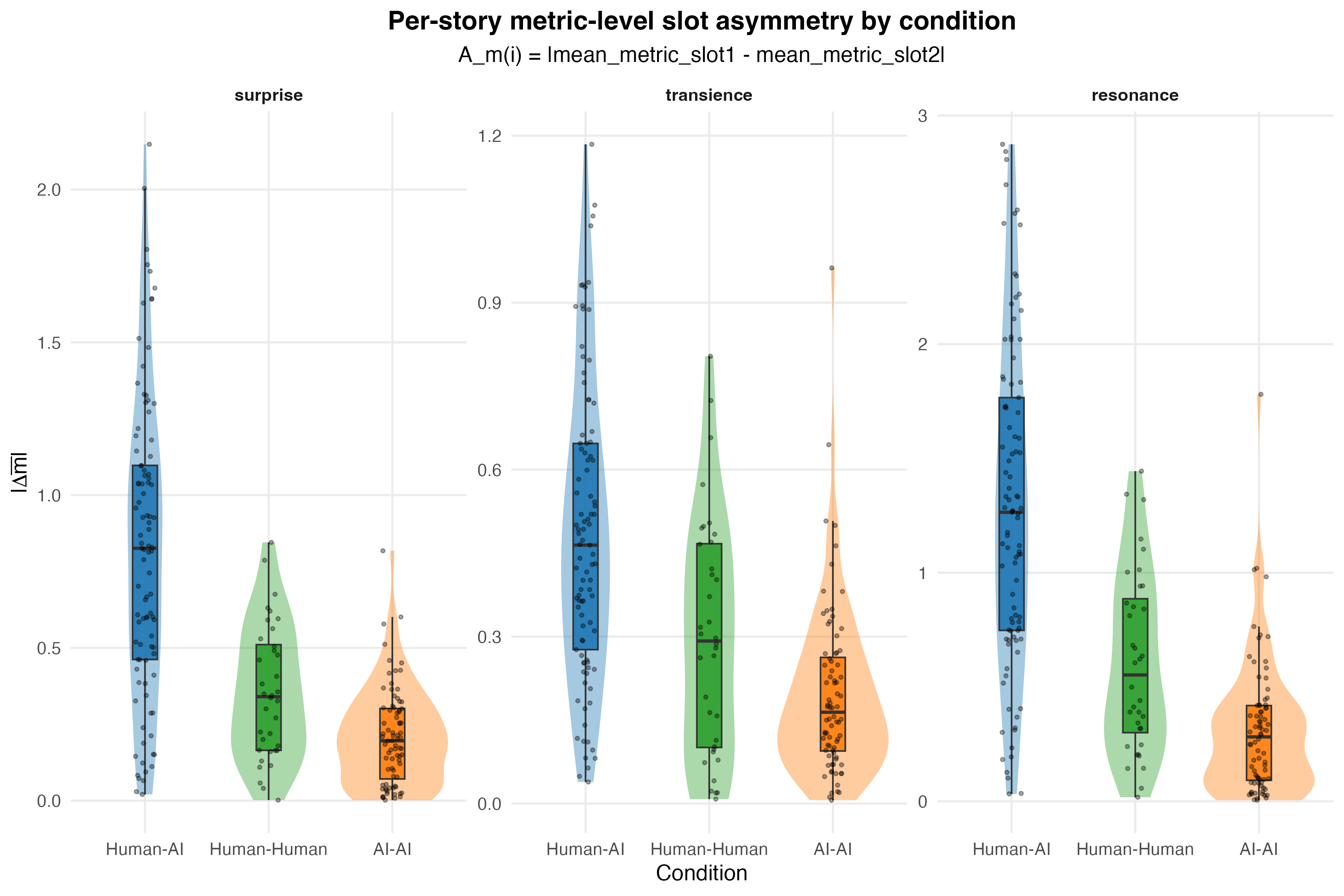}
  \caption{Per-story metric-level slot asymmetry $A_m$ by condition for Novelty, Transience, and Resonance.}
  \label{fig:asymmetry_metrics}
\end{figure}

\begin{figure}[t]
  \centering
  \includegraphics[width=\linewidth]{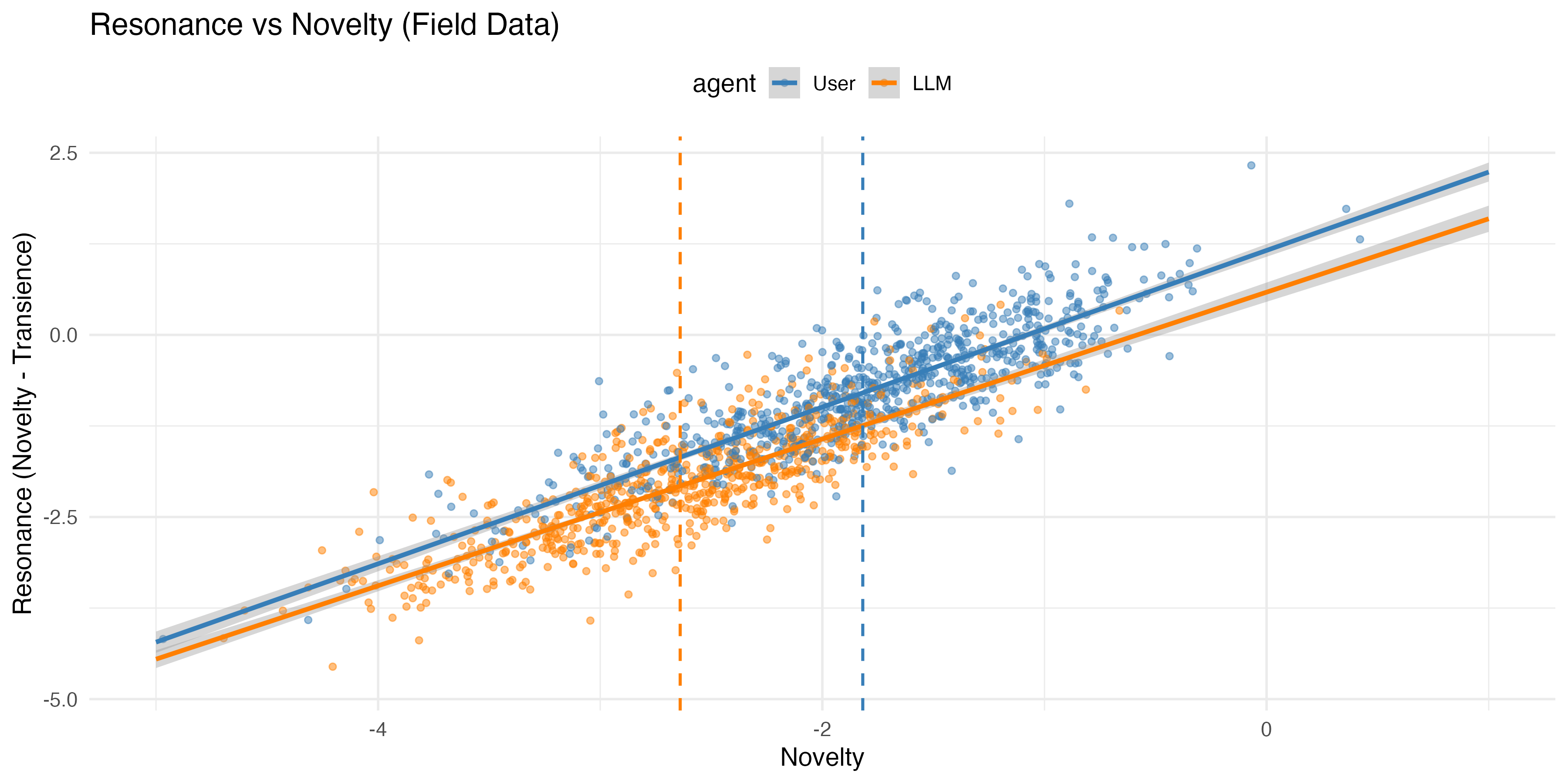}
  \caption{Novelty--resonance relationship within the HA condition.}
  \label{fig:novelty_res-ha}
\end{figure}

\begin{figure}[t]
  \centering
  \includegraphics[width=\linewidth]{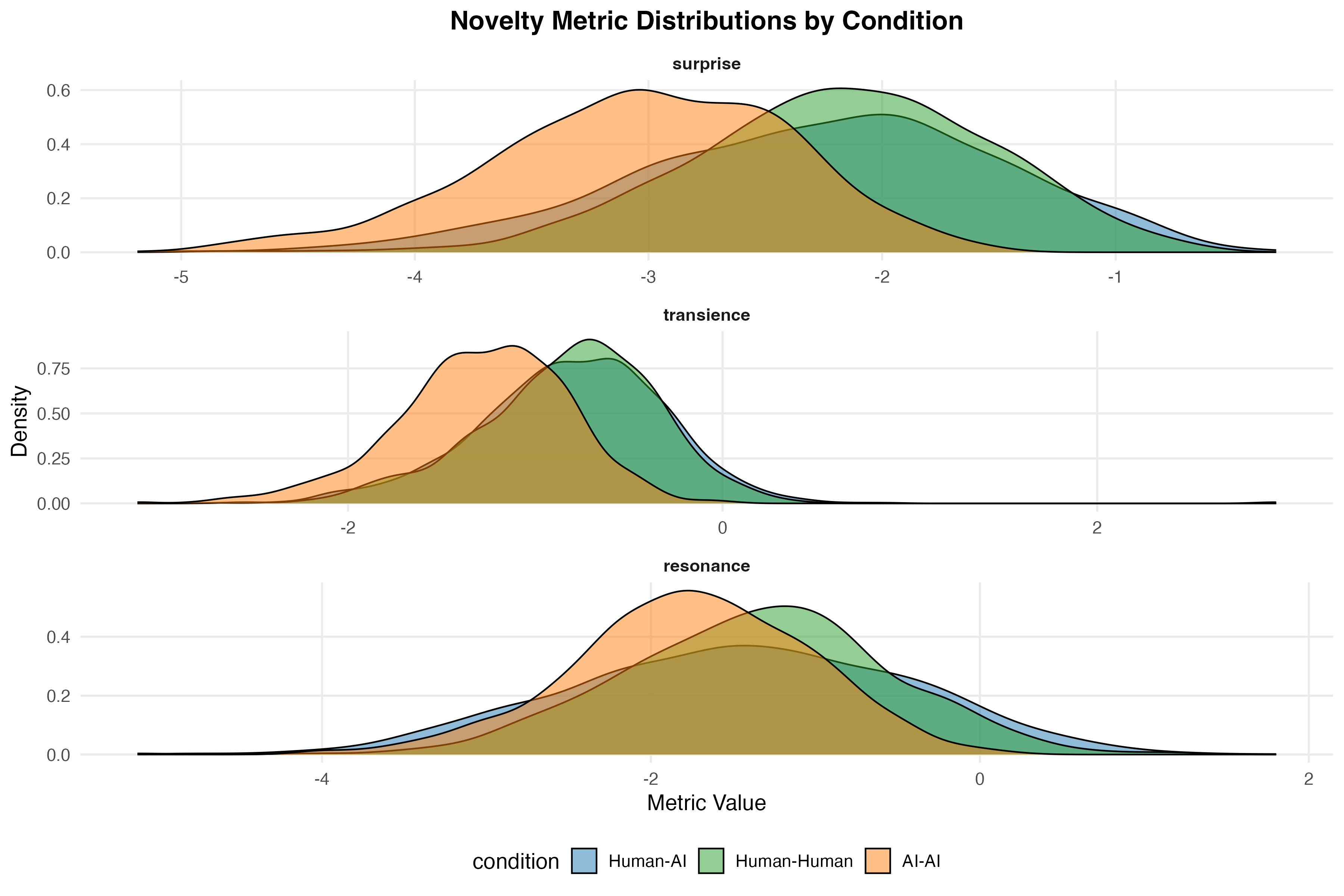}
  \caption{Novelty distributions by condition.}
  \label{fig:novelty_condition_metrics}
\end{figure}

\begin{figure}[t]
  \centering
  \includegraphics[width=\linewidth]{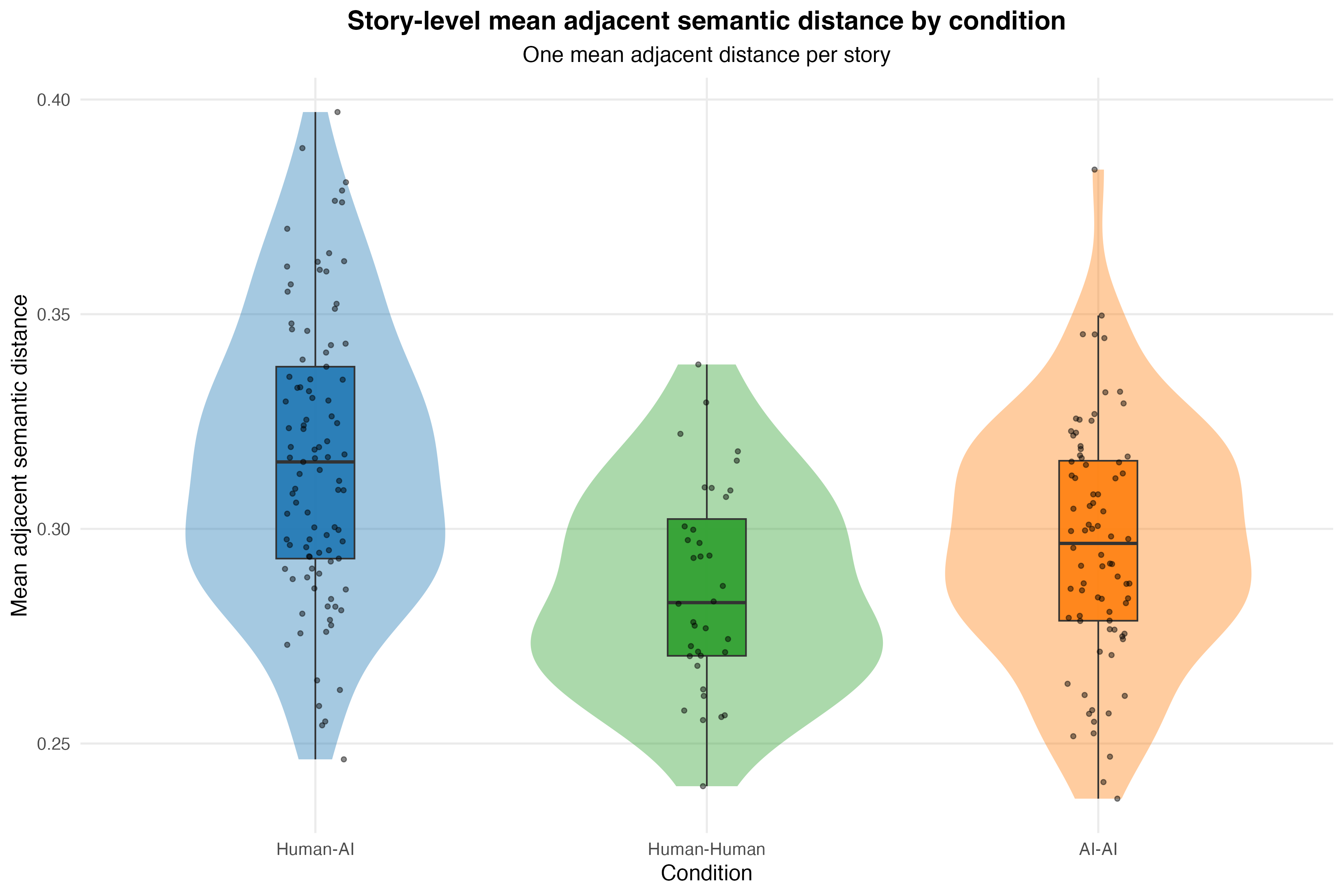}
  \caption{Story-level semantic distance by condition.}
  \label{fig:sem_storylevel}
\end{figure}

Additional diagnostics are shown in Figures~\ref{fig:condition_baseline}, \ref{fig:per_storyasymmetry}, \ref{fig:directional_alignment},  \ref{fig:asymmetry_metrics}, \ref{fig:novelty_res-ha}, \ref{fig:novelty_condition_metrics}, \ref{fig:sem_storylevel}. Valence descriptives are displayed in Table~\ref{tab:valence-descriptives}. Full metric-level asymmetry estimates are reported in Table~\ref{tab:metric-asymmetry-full}.


\begin{table}[t]
\centering\small
\setlength{\tabcolsep}{3.5pt}
\begin{tabular}{@{}lrrrr@{}}
\toprule
Condition & $n$ turns & Mean & SD & Range \\
\midrule
HA & 1700 & $-0.017$ & 0.402 & $[-1.20,\ 1.03]$ \\
HH & 720  & $-0.187$ & 0.389 & $[-1.21,\ 0.89]$ \\
AA & 1599 & $+0.116$ & 0.354 & $[-1.15,\ 1.09]$ \\
\bottomrule
\end{tabular}
\caption{Turn-level projected valence by condition (both author slots
pooled).}
\label{tab:valence-descriptives}
\end{table}

\begin{table*}[!ht]\centering \small \setlength{\tabcolsep}{4pt} \begin{tabular}{llrrcc} \toprule Metric & Condition & $n$ & Mean $A_m$ & 95\% CI & Cliff's $\delta$ vs. HA \\ \midrule Novelty& Human--LLM & 97 & 0.828 & [0.732, 0.926] & -- \\ Novelty& Human--Human & 36 & 0.355 & [0.284, 0.428] & 0.591 \\ Novelty& LLM--LLM & 80 & 0.209 & [0.175, 0.245] & 0.766 \\ \addlinespace Transience & Human--LLM & 97 & 0.484 & [0.432, 0.537] & -- \\ Transience & Human--Human & 36 & 0.300 & [0.233, 0.370] & 0.397 \\ Transience & LLM--LLM & 80 & 0.197 & [0.165, 0.231] & 0.670 \\ \addlinespace Resonance & Human--LLM & 97 & 1.290 & [1.149, 1.432] & -- \\ Resonance & Human--Human & 36 & 0.605 & [0.482, 0.733] & 0.572 \\ Resonance & LLM--LLM & 80 & 0.313 & [0.254, 0.382] & 0.806 \\ \bottomrule \end{tabular} \caption{Metric-level slot asymmetry across conditions. $A_m$ denotes the per-story absolute difference between slot-level mean metric values. Human--LLM shows substantially larger side-level asymmetry in novelty, transience, and resonance than both comparison conditions. Cliff's $\delta$ values compare Human--LLM against each baseline condition.} \label{tab:metric-asymmetry-full} \end{table*}

\end{document}